\documentclass{ws-procs961x669}            
\begin{document}
\title{Baryon Asymmetry and Minimum Length}

\author{Saurya Das$^*$ and Mitja Fridman$^\dagger$}

\address{Theoretical Physics Group and Quantum Alberta, Department of Physics and Astronomy,
University of Lethbridge,
4401 University Drive, Lethbridge,
Alberta, T1K 3M4, Canada\\
$^*$E-mail: saurya.das@uleth.ca\\
$^\dagger$E-mail: fridmanm@uleth.ca}

\author{Gaetano Lambiase}

\address{Dipartimento di Fisica E.R: Caianiello, Universita di Salerno, Via Giovanni Paolo II, 132 - 84084 Fisciano, Salerno, Italy \& INFN - Gruppo Collegato di Salerno, Italy\\
E-mail: lambiase@sa.infn.it}

\author{Elias C. Vagenas}

\address{Theoretical Physics Group, Department of Physics, Kuwait University, P.O. Box 5969, Safat 13060, Kuwait\\
E-mail: elias.vagenas@ku.edu.kw}

\begin{abstract}
We study Quantum Gravity effects in cosmology, and in particular that of the Generalized Uncertainty Principle on the Friedmann equations. We show that the Quantum Gravity induced variations of the energy density and pressure in the radiation dominated era provide a viable explanation of the observed baryon asymmetry in the Universe.
\end{abstract}

\keywords{Quantum Gravity, Quantum Gravity Phenomenology, Minimal Length, Cosmology, Baryon Asymmetry.}

\bodymatter

\section{Introduction}

So far there are two successful theories which describe the world around us. Quantum theory, which describes interactions on the smallest scales on one hand and General Relativity, which describes gravitational interactions on large scales on the other hand \cite{GF,CMW}. There is no straightforward way to combine these two theories to describe physics in a regime, where both contribute, which is why we need a theory of Quantum Gravity (QG). Several QG theories have been proposed, but there is no direct experimental evidence 
yet to verify them. However, phenomenologists have a good intuition what fundamental concepts the regime of QG may imply on physics \cite{CM1,CM2,CM3,CM4,CM5,Buoninfante:2019fwr,Blasone:2019wad,CM6,CM7}.
One such concept is the existence of a minimal measurable length, predicted my all or most models of QG. 
A natural way to incorporate minimal length in our considerations is to modify the Heisenberg uncertainty principle \cite{GUP1,GUP2,GUP3,GUP4,KMM,FC,Das:2009hs, Ali:2010yn,ADV0,IPK,Scardigli:2014qka,SLV,KP,VenezGrossMende,Basilakos:2010vs, Das:2020ujn,CM8GS,Iorio:2019wtn,luciano}. The modified Heisenberg uncertainty principle is called the Generalized Uncertainty Principle (GUP) and in the commutator form it reads as \cite{ADV0}
\begin{equation}
    \label{gup}
     [x_i,p_j]=i\hbar\left(\delta_{ij}-\alpha\left(p\delta_{ij}+\frac{p_ip_j}{p}\right)+\beta\left(p^2\delta_{ij}+3p_ip_j\right)\right)
\end{equation}
where $\alpha \equiv \alpha_0/(M_{P}c)$, 
$\beta \equiv \beta_0/(M_{P}c)^2$, 
$\alpha_0$, $\beta_0$ are the dimensionless
linear and quadratic GUP parameters and $M_{P}=\sqrt{\hbar c/G}$ 
is the Planck mass. This phenomenological model describes a length scale $\alpha_0\, \ell_P$ and $\sqrt{\beta_0}\,\ell_P$ between the electroweak scale, i.e., $\ell_{EW}\approx10^{-18}\,\mathrm{m}$, and the Planck scale, i.e., $\ell_P=\sqrt{\hbar G/c^3}\approx10^{-35}\,\mathrm{m}$. We apply the GUP to quantum quantum mechanical and gravitational systems and predict deviations from standard theory, which we hope experiments can detect now or in the future, or provide explanations for observed phenomena, as in the case of this work, namely the observed baryon asymmetry that is observed in the Universe.

The asymmetry between matter and anti-matter has been an unsolved problem in physics for a long time. Standard theories such as quantum theory and general relativity suggest that there is no reason why the asymmetry between matter and anti-matter in the Universe should exist \cite{CDS}. 
This means there should be an equal amount of matter as there is anti-matter in the Universe, while observations show a very large asymmetry. To explain this observed asymmetry we have to satisfy the so-called Sakharov conditions \cite{Sakh}:
\begin{enumerate}
    \item[1)] violation of baryon number $B$,
    \item[2)] violation of C and CP symmetries and
    \item[3)] deviation from thermal equilibrium.
\end{enumerate}
If a proposed mechanism is able to satisfy the above three Sakharov conditions, then it may be possible to explain the origin of the observed baryon asymmetry in the Universe. The mechanism we use is an interaction term from supergravity theories which couples the baryon current and space-time to satisfy the first two Sakharov conditions, and the holographic principle with the thermodynamics of horizons to derive the GUP-modified Friedmann equations, which satisfy the third Sakharov condition. Several proposals how GUP modifies the Friedmann equations have been presented so far \cite{Cai:2008ys,Zhu:2008cg,AA,Giardino:2020myz}. In this way we provide a viable explanation of the observed baryon symmetry in the Universe using the GUP. 

The paper is structured as follows. In section \ref{sec2} we modify the Bekenstein-Hawking entropy using GUP, then in section \ref{sec3}, by using the holographic principle we modify the Friedmann equations, and in section \ref{sec4} by using these modified Friedmann equations we attempt to explain the observed baryon symmetry in the Universe. In section \ref{sec5} we summarize the work with concluding remarks.

\section{Modified Bekenstein-Hawking entropy}
\label{sec2}

In order to modify the Friedmann equations, using the holographic principle, we need to modify the Bekenstein-Hawking entropy formula, since the Friedmann equations can be derived from the first law of thermodynamics, which requires the change in entropy. The holographic principle makes it possible to write a $d$-dimensional theory on a $d-1$ dimensional boundary \cite{GTH,LS}. We can re-write the GUP from Eq. (\ref{gup}) in the usual form \cite{GUPAM} (where $\langle p\rangle=0$)
\begin{equation}
\label{upg}
    \Delta x\Delta p\gtrsim\left[1-\alpha\Delta p+4\beta\Delta p^2\right]~,
\end{equation}
where we have set the numerical prefactor to be $\mathcal{O}(1)$ \cite{MV,ACAP,LJG}.

When a particle with energy $E$ gets absorbed by a horizon, the horizon area changes by $\Delta A_{min}\geq8\pi\ell_P^2E\,\Delta x$, where $\Delta x$ is the uncertainty in position of the absorbed particle \cite{DC,CR}. To obtain the energy of the particle $E$, we solve the quadratic equation for $\Delta p$ from Eq. (\ref{upg}) and obtain the expression
\begin{eqnarray}
\label{delp}
    E=\Delta p\gtrsim\frac{\Delta x+\alpha}{8\beta}\left(1-\sqrt{1-\frac{16\beta}{\Delta x^2+2\alpha\Delta x+\alpha^2}}\right),
\end{eqnarray}

where we chose the negative solution, since we want to obtain the smallest change in area of the apparent horizon and this solution reduces to the standard Heisenberg uncertainty for $\alpha,\beta\longrightarrow0$. The position uncertainty $\Delta x$ is the diameter of the observable Universe $\Delta x =2r_S$, where $r_S$ is the Schwarzschild radius. The square of $\Delta x$ is then related to the area of the apparent cosmic horizon as $\Delta x^2=A/\pi$. Plugging all of the above in the change in $\Delta A_{min}$, we get
\begin{eqnarray}
    \Delta A_{min}\simeq\lambda\frac{\ell_P^2(A+\alpha\sqrt{\pi}A^{1/2})}{\beta}\left(1-\sqrt{1-\frac{16\pi\beta}{A+2\alpha\sqrt{\pi}A^{1/2}+\alpha^2\pi}}\right),
\end{eqnarray}
where $\lambda$ is determined by the Bekenstein-Hawking entropy formula 
$b/\lambda=2\pi$. Here 
$b=\Delta S_{min}=\ln{2}$ is the minimal increase in entropy, corresponding to one bit of information \cite{AA}.

The Bekenstein-Hawking entropy was first introduced for a black hole, but the holographic principle suggests that it can be used for any horizon. Therefore we can write it for a cosmic horizon in the same way \cite{JB,SH}
\begin{eqnarray}
\label{bhe}
S=\frac{A}{4G}=\frac{A}{4\ell_P^2}~.
\end{eqnarray}
If we introduce some modification to the Heisenberg uncertainty principle or we modify the entropy in any other way, the only variable in this equation, the area, gets modified \cite{Cai:2008ys}. The Bekenstein-Hawking entropy then reads $S=\tfrac{f(A)}{4\ell_P^2}$. Since the first law of thermodynamics includes a differential of entropy, it useful to calculate its derivative
\begin{eqnarray}
\label{gdventropy}
    \frac{\mathrm{d}S}{\mathrm{d}A}=\frac{f'(A)}{4\ell_P^2}~.
\end{eqnarray}

On the other hand, the above derivative can be written using the minimal change in entropy $\Delta S_{min}$ and $\Delta A_{min}$
\begin{eqnarray}
\label{dventropy}
    \frac{\mathrm{d}S}{\mathrm{d}A}=\frac{\Delta S_{min}}{\Delta A_{min}}=\frac{\beta^*}{8\ell_P^2\left(A+\alpha^*A^{1/2}-\sqrt{A^2+2\alpha^*A^{3/2}+(\alpha^{*2}-\beta^*)A}\right)}~,
\end{eqnarray}
where  
$\alpha^*=\sqrt{\pi}\alpha$ and $\beta^*=16\pi\beta$.
By comparing Eqs. (\ref{gdventropy}) and (\ref{dventropy}) we can read out what $f'(A)$ is
\begin{eqnarray}
\label{dfa}
    f'(A)=\frac{1}{2}\frac{\beta^*}{\left(A+\alpha^*A^{1/2}-\sqrt{A^2+2\alpha^*A^{3/2}+(\alpha^{*2}-\beta^*)A}\right)}~.
\end{eqnarray}
To obtain the modified Bekenstein-Hawking entropy we must integrate Eq. (\ref{gdventropy}) over area $A$, from $A$ to $\infty$, using $f'(A)$ from Eq. (\ref{dfa}). The exact GUP-modified Bekenstein-Hawking entropy then reads
\begin{eqnarray}
S &=& \frac{1}{8\ell_P^2}\left[A\left(1+\sqrt{1+2\alpha^*\frac{1}{A^{1/2}}+(\alpha^{*2}-\beta^*)\frac{1}{A}}\right)\right. \nonumber \\
&+&\left.\alpha^*A^{1/2}\left(2+\sqrt{1+2\alpha^*\frac{1}{A^{1/2}}+(\alpha^{*2}-\beta^*)\frac{1}{A}}\right)\right. \nonumber \\
&-&\!\!\!\left.\beta^*\ln{\left(1+\frac{A^{1/2}}{\alpha^*}\left(1+\sqrt{1+2\alpha^*\frac{1}{A^{1/2}}+(\alpha^{*2}-\beta^*)\frac{1}{A}}\right)\right)}\right]~,
\end{eqnarray}
where we obtain the standard result from Eq.(\ref{bhe}) for $\alpha^*,\beta^*\longrightarrow0$.

\section{Modified Friedmann equations}
\label{sec3}

The Friedmann equations describe the evolution of a dynamical Universe and provide the foundations for the standard model of cosmology. They can be derived in different ways. The most common is by solving the Einstein equations in $(n+1)$-dimensional space-time for a FLRW metric
\begin{eqnarray}
\label{metric}
\mathrm{d}s^2=h_{cd}\mathrm{d}x^c\mathrm{d}x^d+\tilde{r}^2\mathrm{d}\Omega_{n-1}^2~,
\end{eqnarray}
where $x^c=(t,r)$, $\tilde{r}=a(t)\,r$, $h_{cd}={\text{diag}}(-1,a^2/(1-kr^2))$,
$\mathrm{d}\Omega_{n-1}$ is the angular part of the $(n-1)$-dimensional sphere, $a=a(t)$ is the scale factor, $r$ is the comoving radius and $k$ is the spatial curvature constant. Indices $c$ and $d$ only take values $0$ and $1$. The other way to derive the Friedmann equations is using the holographic principle, which takes advantage of the thermodynamics of horizons. 

To generalize
the Friedmann equations to include minimum length effects we derive the same from the first law of thermodynamics, with the difference of using the modified Bekenstein-Hawking entropy, derived in the previous section. The first law of thermodynamics for the content inside an apparent horizon is given by  \cite{Cai:2008ys,Zhu:2008cg,AA,Giardino:2020myz}
\begin{eqnarray}
\label{flt}
\mathrm{d}E=T\mathrm{d}S+W\mathrm{d}V~,
\end{eqnarray}
where $E$ is the energy contained inside the apparent horizon, given by $E=\rho V$,  $V$ is the volume of an $n$-dimensional sphere, given by $V=\Omega_n\,\tilde{r}_A^{\,n}$, with $\Omega_n=\tfrac{\pi^{n/2}}{\Gamma(n/2+1)}$ which has an area of $A=n\, \Omega_{n}\,\tilde{r}_{A}^{\,n-1}$), $T$ is the Hawking temperature given by \cite{CK,Cai:2008gw}
\begin{eqnarray}
\label{htemp}
T=\frac{\kappa}{2\pi}=-\frac{1}{2\pi\tilde{r}_A}\left(1-\frac{\dot{\tilde{r}}_A}{2H\tilde{r}_A}\right)~,
\end{eqnarray}
$\mathrm{d}S$ is the differential of the Bekenstein-Hawking entropy given by Eq. (\ref{gdventropy}), and  $W$ is the work density which reads as \cite{SAH}

\begin{eqnarray}
\label{wdens}
W=-\frac{1}{2}T^{cd}h_{cd}=\frac{1}{2}\left(\rho-p\right)~.
\end{eqnarray}
In the above $T^{cd}$ is a projection of the energy-momentum tensor $T_{\mu\nu}=\left(\rho+p\right)u_\mu u_\nu+{p}g_{\mu\nu}$ on the $t-r$ subspace.
What remains is to differentiate the energy and volume, write everything in terms of $\mathrm{d}A$, and plug the above equations in the first law of thermodynamics from Eq. (\ref{flt}). From here we derive the first Friedmann equation
\begin{eqnarray}
\label{fe1}
    -\frac{8\pi G}{n-1}\left(\rho+p\right)=\left(\dot{H}-\frac{k}{a^2}\right)f'(A)~.
\end{eqnarray}
By using the continuity equation $\dot{\rho}+nH\left(\rho+p\right)=0$, originating from the conservation of energy $T^{\mu\nu}_{\,\,\,\,\,\,;\nu}=0$, with Eq.(\ref{fe1}) and integrating it, one obtains the second Friedmann equation

\begin{eqnarray}
\label{fe2}
    -\frac{8\pi G}{n(n-1)}\rho=\frac{\left(n\Omega_n\right)^{\frac{n+1}{n-1}}}{n(n-1)\Omega_n}\int f'(A)\frac{\mathrm{d}A}{A^{\frac{n+1}{n-1}}}~.
\end{eqnarray}
By setting $f'(A)=1$ ($f(A)=A$ in standard theory) we obtain the standard Friedmann equations as we expect. However, to obtain GUP corrections to the Friedmann equations, we plug $f'(A)$ from Eq. (\ref{dfa}) in Eqs. (\ref{fe1}) and (\ref{fe2}). Both of the Friedmann equations have exact solutions in $n=3$, which read as
\begin{eqnarray}
\label{emfe1}
&-&4\pi G\left(\rho+p\right)=\left(\dot{H}-\frac{k}{a^2}\right)  \\
&\times& \frac{\beta^*}{8\pi}\frac{\left(H^2+\frac{k}{a^2}\right)}{1+\frac{\alpha^*}{(4\pi)^{1/2}}\left(H^2+\frac{k}{a^2}\right)^{1/2}-\sqrt{1+\frac{2\alpha^*}{(4\pi)^{1/2}}\left(H^2+\frac{k}{a^2}\right)^{1/2}+\frac{(\alpha^{*2}-\beta^*)}{4\pi}\left(H^2+\frac{k}{a^2}\right)}}~ \nonumber
\end{eqnarray}
and
\begin{eqnarray}
\label{emfe2}
\frac{8\pi G}{3}(\rho-\Lambda ) &=& \frac{1}{2}\left(H^2+\frac{k}{a^2}\right)+\frac{\alpha^*}{3(4\pi)^{1/2}}\left(H^2+\frac{k}{a^2}\right)^{3/2}+\frac{2\pi (\alpha^{*2}+2\beta^*)}{3(\alpha^{*2}-\beta^*)^2}\nonumber \\
&+&\left[\frac{1}{3}\left(H^2+\frac{k}{a^2}\right)+\frac{(4\pi)^{1/2} \alpha^*}{6(\alpha^{*2}-\beta^*)}\left(H^2+\frac{k}{a^2}\right)^{1/2}-\frac{2\pi(\alpha^{*2}+2\beta^*)}{3(\alpha^{*2}-\beta^*)^2}\right] \nonumber \\
 &\times& \sqrt{1+\frac{2\alpha^*}{(4\pi)^{1/2}}\left(H^2+\frac{k}{a^2}\right)^{\!\!1/2}\!\!\!\!\!\!\!+\frac{(\alpha^{*2}-\beta^*)}{4\pi}\left(H^2+\frac{k}{a^2}\right)} \nonumber \\
&+& \frac{2\pi\alpha^*\beta^*}{(\alpha^{*2}-\beta^*)^{5/2}}\ln{\left[1+\frac{(\alpha^{*2}-\beta^*)}{(4\pi)^{1/2}(\alpha^*+\sqrt{\alpha^{*2}-\beta^*})}\left(H^2+\frac{k}{a^2}\right)^{1/2}\right.} \nonumber \\
&+&\left.\frac{\sqrt{\alpha^{*2}-\beta^*}}{\alpha^*+\sqrt{\alpha^{*2}-\beta^*}}\right. \nonumber \\
&\times&\!\!\left.\left(\sqrt{1+\frac{2\alpha^*}{(4\pi)^{1/2}}\left(H^2+\frac{k}{a^2}\right)^{\!\!1/2}\!\!\!\!\!\!\!+\frac{(\alpha^{*2}-\beta^*)}{4\pi}\left(H^2+\frac{k}{a^2}\right)}-1\right)\right], 
\end{eqnarray}
where
\begin{eqnarray}
A=4\pi \tilde{r}_A^2=\frac{4\pi}{H^2+\frac{k}{a^2}}~
\end{eqnarray}
was used. We are interested in the radiation dominated era, since the baryon asymmetry was frozen in at that epoch. In that epoch the small cosmological constant $\Lambda$ can be ignored, since $\rho\gg\Lambda$ and the curvature constant can be set to $k=0$, which is consistent with the observations which show a flat Universe. The modified Friedmann equations then get simplified to
\begin{eqnarray}
\label{cmfe1}
-4\pi G\left(\rho+p\right)=\frac{\beta^*\dot{H}}{8\pi}\frac{H^2}{1+\frac{\alpha^*}{(4\pi)^{1/2}}H-\sqrt{1+\frac{2\alpha^*}{(4\pi)^{1/2}}H+\frac{(\alpha^{*2}-\beta^*)}{4\pi}H^2}}~
\end{eqnarray}
and
\begin{eqnarray}
\label{cmfe2}
\frac{8\pi G}{3}\rho&=&\frac{1}{2}H^2+\frac{\alpha^*}{3(4\pi)^{1/2}}H^3+\frac{2\pi(\alpha^{*2}+2\beta^*)}{3(\alpha^{*2}-\beta^*)^2}\nonumber \\
&+&\left[\frac{1}{3}H^2+\frac{(4\pi)^{1/2}\alpha^*}{6(\alpha^{*2}-\beta^*)}H-\frac{2\pi(\alpha^{*2}+2\beta^*)}{3(\alpha^{*2}-\beta^*)^2}\right]\sqrt{1+\frac{2\alpha^*}{(4\pi)^{1/2}}H+\frac{(\alpha^{*2}-\beta^*)}{4\pi}H^2} \nonumber \\
&+&\frac{2\pi\alpha^*\beta^*}{(\alpha^{*2}-\beta^*)^{5/2}}\ln{\left[1+\frac{(\alpha^{*2}-\beta^*)}{(4\pi)^{1/2}(\alpha^*+\sqrt{\alpha^{*2}-\beta^*})}H\right.} \nonumber \\
&+&{\left.\frac{\sqrt{\alpha^{*2}-\beta^*}}{\alpha^*+\sqrt{\alpha^{*2}-\beta^*}}\left(\sqrt{1+\frac{2\alpha^*}{(4\pi)^{1/2}}H+\frac{(\alpha^{*2}-\beta^*)}{4\pi}H^2}-1\right)\right]}~.
\end{eqnarray}
The Taylor expansions of the above modified Friedmann equations up to first order in $\alpha^*$ and $\beta^*$ turn out to be
\begin{eqnarray}
\label{tfe1}
\displaystyle{-4\pi G\left(\rho+p\right)=\dot{H}\left(1+\frac{\alpha^*}{2\sqrt{\pi}}H-\frac{\beta^*}{16\pi}H^2\right)}
\end{eqnarray}
and
\begin{eqnarray}
\label{tfe2}
           \displaystyle{\frac{8\pi G}{3}\rho=H^2\left(1+\frac{\alpha^*}{3\sqrt{\pi}}H-\frac{\beta^*}{32\pi}H^2\right)}~.
\end{eqnarray}
The above simplified equations are used to study the breaking of thermal equilibrium in the radiation dominated era, which satisfies one of the Sakharov conditions.

\section{Gravitational Baryogenesis}
\label{sec4}

To provide an explanation for the observed baryon asymmetry in the Universe, all Sakharov conditions must be satisfied. Theories of supergravity introduce an extra interaction term to the action which couples the baryon current and space-time \cite{gravbar} 
\begin{eqnarray}
\label{intterm}
\frac{1}{M_*^{2}}\int\mathrm{d}^4x\sqrt{-g}\,J^\mu\partial_\mu R\,,
\end{eqnarray}
and thus violates the C, CP and CPT symmetries. In the above $M_*$ is the cutoff scale characterizing the effective theory \cite{allGB}. The magnitude of $M_*$ is chosen to be in the regime of the Grand Unified Theories (GUT), where interactions which violate the baryon number exist. The term in Eq. (\ref{intterm}) therefore satisfies the first two Sakharov conditions. We can see that the integrand from Eq. (\ref{intterm}) reduces to
\begin{eqnarray}
\label{integrand}
\frac{1}{M_*^2}J^\mu\partial_\mu R=\frac{1}{M_*^2}(n_B-n_{\Bar{B}})\dot{R}~,
\end{eqnarray}
where $n_B$ ($n_{\Bar{B}}$) is the baryon (anti-baryon) number density. In the above equation only the time derivative survives, since the spatial derivative of the Ricci scalar vanishes $\nabla R=0$ for the FLRW metric from Eq. (\ref{metric}). Because the integrand from Eq. (\ref{integrand}) corresponds to the energy density term of the grand canonical ensemble, we can read the chemical potential for baryons and anti-baryons from it as
\begin{eqnarray}
\label{chempot}
\mu_B=-\mu_{\Bar{B}}=-\frac{\dot{R}}{M_*^2}~.
\end{eqnarray}

To study the baryon asymmetry, we need a suitable variable, which parametrizes this asymmetry. Following  Ref. \cite{KT}, the baryon asymmetry parameter is defined as 
\begin{eqnarray}
\label{baspar}
\eta\equiv\frac{n_B-n_{\Bar{B}}}{n_\gamma}\simeq7\frac{n_B-n_{\Bar{B}}}{s}~,
\end{eqnarray}
where the photon number density $n_\gamma$ and the entropy density $s$ are related by a constant factor $n_\gamma\simeq s/7$ since the epoch of $e^\pm$ annihilation. For relativistic particles, the entropy density turns out to be \cite{KT}
\begin{eqnarray}
\label{entdens}
s=\frac{2\pi^2g_{*}}{45}T^3
\end{eqnarray}
and the baryon number density \cite{KT}
\begin{eqnarray}
\label{numdens}
n_B-n_{\Bar{B}}=\frac{g_b}{6}\mu_BT^2~.
\end{eqnarray}
In the above equations $g_{*}\approx106$ is the number of degrees of freedom for particles which contribute to the entropy of the universe and $g_b\sim\mathcal{O}(1)$ is the number of intrinsic degrees of freedom for baryons \cite{KT}. Plugging Eqs. (\ref{chempot}), (\ref{entdens}) and (\ref{numdens}) in Eq. (\ref{baspar}) the baryon asymmetry parameter reads
\begin{eqnarray}
\label{fbas}
\eta\simeq-\frac{105\,g_b}{4\pi^2g_*}\frac{\dot{R}}{M_*^2T}\bigg|_{T_D}~.
\end{eqnarray}
The above parameter is evaluated at some decoupling temperature $T_D$. This decoupling temperature represents the energy scale at which the baryon violating interactions freeze out and go out of equilibrium. The baryon number is conserved below this temperature for all future epochs. 

Next, we use the modified Friedmann equations which turn out to break thermal equilibrium. 
To see that they indeed break thermal equilibrium, we write the energy density and pressure as their equilibrium values plus some small variations
\begin{eqnarray}
\label{perturb}
\rho=\rho_0+\delta\rho\,,\qquad p=p_0+\delta p~.
\end{eqnarray}
If there is no variations $\delta\rho=\delta p=0$, we have thermal equilibrium and no baryon asymmetry is produced due to the third Sakharov condition. We require some variations $\delta\rho,\delta p\neq0$ to deviate from the thermal equilibrium. To obtain non-vanishing variations we plug the energy density and pressure from Eq. (\ref{perturb}) into the modified Friedman equations from Eqs. (\ref{tfe1}) and (\ref{tfe2}) to express $\delta\rho$ and $\delta p$ in terms of the GUP parameters and the equation of state $p_0=w\rho_0$. The variations then read
\begin{eqnarray}
\label{drho}
\delta\rho=\frac{\alpha^*}{3}\sqrt{\frac{8G}{3}}\rho_0^{3/2}-\frac{\beta^*}{12}G\rho_0^2
\end{eqnarray}
and
\begin{eqnarray}
\label{dp}
\delta p=\frac{\alpha^*}{6}\sqrt{\frac{8G}{3}}(1+3w)\rho_0^{3/2}-\frac{\beta^*}{12}G(1+2w)\rho_0^2~,
\end{eqnarray}
We can  see if there is no GUP corrections $\alpha^*,\beta^*=0$, we get no variations in the energy density and pressure and thermal equilibrium is preserved.

The trace of the Einstein equations give us the Ricci scalar $R=-8\pi G\,T_g=-8\pi G\,(\rho-3p)$, where we plug in the energy density and pressure from Eq. (\ref{perturb}) with their respective variations from Eqs. (\ref{drho}) and (\ref{dp}). We compute its time derivative, use the continuity equation for $n=3$ and evaluate the whole expression in the radiation-dominated era with $w=1/3$, where the baryon violating interactions froze out. The time derivative of the Ricci scalar then reads
\begin{eqnarray}
\label{dtricci}
\dot{R}=-\alpha^*\frac{256}{3}\,\pi^{3/2}\,G^2\rho_0^2+\beta^*\,8\left(\frac{8\pi}{3}\right)^{3/2}\!\!\!G^{5/2}\rho_0^{5/2}~.
\end{eqnarray}
We can see that the above derivative would also vanish if there would be no GUP corrections because every term is proportional to $\alpha^*$ or $\beta^*$. We plug Eq. (\ref{dtricci}) in Eq. (\ref{fbas}) and use the equilibrium energy density for relativistic particles \cite{KT}
\begin{eqnarray}
\rho_0=\frac{\pi g_*}{30}T^4~,
\end{eqnarray}
to obtain the expression for the predicted baryon asymmetry
\begin{eqnarray}
\label{bas}
\eta=\alpha_0\frac{112\,\pi^2g_*g_b}{45}\left(\frac{T_D}{M_P}\right)^7\left(\frac{M_P}{M_*}\right)^2-\beta_0\frac{896\sqrt{5}\,\pi^3g_*^{3/2}g_b}{675}\left(\frac{T_D}{M_P}\right)^9\left(\frac{M_P}{M_*}\right)^2~,
\end{eqnarray}
where we expressed the gravitational constant in terms of the Planck mass, i.e., $M_P$, as 
$G=1/M_P^2$. From the above we can see that the baryon asymmetry is dependent on $\alpha_0$ and $\beta_0$ parameters and the scales $M_*$ and $T_D$.

To compare Eq. (\ref{bas}) to the observed baryon asymmetry we plug in the suggested values for $M_*=M_P/\sqrt{8\pi}$, the reducep Planck mass and $T_D=M_I\sim 2\times10^{16}\,\mathrm{GeV}$ which corresponds to the tensor mode fluctuations in the inflationary scale \cite{Kinney:2006qm}. We then obtain
\begin{eqnarray}
\label{basnum}
\eta=\alpha_0\,2.08\times10^{-15}-\beta_0\,2.16\times10^{-19}~,
\end{eqnarray}
where we can see the baryon asymmetry vanishes for $\alpha_0,\beta_0\longrightarrow0$. If $\alpha_0,\beta_0\neq0$ the baryon asymmetry is explained by the existence of a minimal length scale.

We compare the predicted baryon asymmetry from Eq. (\ref{basnum}) to the observational bounds $5.7\times10^{-11}\lesssim\eta\lesssim9.9\times10^{-11}$ \cite{baryogenesis}. By doing this we can fix the GUP parameters to their actual values for different models:
\begin{itemize}
    \item Only linear GUP ($\beta_0=0$): $2.74\times10^4\lesssim\alpha_0\lesssim4.76\times10^4$
    \item Only quadratic GUP ($\alpha_0=0$): $-4.58\times10^8\lesssim\beta_0\lesssim-2.64\times10^8$
    \item Linear and Quadratic GUP ($\beta_0=-\alpha_0^2$): $1.21\times10^4\lesssim\alpha_0\lesssim1.71\times10^4$ and $-2.92\times10^8\lesssim\beta_0\lesssim-1.46\times10^8$
    \item Linear and Quadratic GUP: $\alpha_0\gtrsim4.81\times10^3$ and $\beta_0\lesssim-1.48\times10^8$
\end{itemize}
We can see that the GUP parameters should have values $\alpha_0\approx10^4$ and $\beta_0\approx-10^8$, which suggest a new length scale of $\ell_{new}=\alpha_0\ell_P\sim\sqrt{-\beta_0}\ell_P\approx10^{-31}\,\mathrm{m}$.

\section{Conclusion}
\label{sec5}

With this work we test the fundamental physics of nature. More specifically, how the minimal length can affect phenomena on large scales such as in this case, the baryon asymmetry in the Universe, which is still a problem in physics up to this day. The mechanism which we propose satisfies all Sakharov conditions and thus offers a viable explanation of the observed baryon asymmetry in the Universe. We used an interaction term which couples the baryon current and space-time to satisfy the first two Sakharov conditions, and using the holographic principle we modified the Bekenstein-Hawking entropy and the Friedmann equations to break thermal equilibrium and thus satisfying the third and last Sakharov condition. The modified Friedmann equations can also be used more broadly in cosmology and can provide a very rich phenomenology. 

Our findings suggest that minimal length effects could have been the cause for generating the observed baryon symmetry. Our work does not constrain the GUP parameters to an improved upper bound because, but rather fixes the values of the GUP parameters $\alpha_0$ and $\beta_0$ and thus sets the new length scale to be $\ell_{new}=\alpha_0\ell_P\sim\sqrt{-\beta_0}\ell_P\approx10^{-31}\,\mathrm{m}$.

We noticed that the quadratic GUP parameter turns out to be negative. This is consistent with other cosmological studies using quadratic GUP \cite{Nenmeli:2021orl}, non-trivial structures of space-time \cite{Buoninfante:2019fwr,JKS,Nenmeli:2021orl,KMM,vilasi} and a crystal lattice  considerations with GUP \cite{JKS}. This can suggest that space-time can have a granular structure at the Planck scale \cite{Ali:2009zq,Das:2010zf,Das:2020ujn,Deb:2016psq}.

\end{document}